\documentclass[conference]{IEEEtran}
\IEEEoverridecommandlockouts
\usepackage{cite}
\usepackage{amsmath,amssymb,amsfonts}
\usepackage{algorithmic}
\usepackage{graphicx}
\usepackage{textcomp}
\usepackage{xcolor}
\usepackage{fancyhdr}
\usepackage{booktabs}
\usepackage{url}
\usepackage{amsmath,amssymb,amsfonts}
\usepackage{tabularx}
\usepackage{amsthm}
\usepackage{multirow}
\usepackage{multicol}
\usepackage{caption}
\usepackage{tikz}
\usetikzlibrary{quantikz2}
\usepackage{pgfplots}
\usepackage{subcaption}
\pgfplotsset{compat=1.18}
\def\BibTeX{{\rm B\kern-.05em{\sc i\kern-.025em b}\kern-.08em
    T\kern-.1667em\lower.7ex\hbox{E}\kern-.125emX}}
    
\newtheorem{definition}{Definition}

\begin{document}
\title{StabilizerBench: A Benchmark for AI-Assisted\\Quantum Error Correction Circuit Synthesis}

\author{
 Andres Paz$^{1}$, Christian Tarta$^{1}$, Cordelia Yuqiao Li$^{1}$, Mayee Sun$^{1}$, Sarju Patel$^{1}$, and Sylvie Lausier$^{1}$
 
\thanks{$^{1}$University of Washington, Seattle, WA, USA}
\thanks{Corresponding authors: anpaz@cs.washington.edu, yuqiaoli@uw.edu}

}

\maketitle

\begin{abstract}
As quantum hardware scales toward fault-tolerant operation, the demand for correct quantum error correction (QEC) circuits far outpaces manual design capacity.
AI agents offer a promising path to automating this synthesis, yet no benchmark exists to measure their progress on the specialized task of generating QEC circuits.
We introduce \textsc{StabilizerBench}, a benchmark suite of 192 stabilizer codes spanning 14~families, 4--196~qubits, and distances 2–21, organized into three tasks of increasing difficulty: state-preparation circuit generation, circuit optimization under semantic constraints, and fault-tolerant circuit synthesis.
Although motivated by QEC, stabilizer circuits exercise the same core competencies required for general quantum programming, including gate decomposition, qubit routing, and semantic-preserving transformations, while admitting efficient verification via the Gottesman--Knill theorem, enabling the benchmark to scale to large codes without the exponential cost of full unitary comparison.
We define a unified, generator-weighted scoring system with two tiers: a capability score that measures breadth of success and a quality score that captures circuit merit. We also introduce novel continuous fault-tolerance and optimization metrics that grade error resilience beyond binary pass/fail.
Following the design of classical benchmarks such as SWE-bench, \textsc{StabilizerBench} specifies inputs, verification oracles, and scoring but leaves prompts and agent strategies open, ensuring the benchmark remains durable as techniques evolve.
We validate the benchmark's design by evaluating three frontier AI agents, confirming that it is discriminative across models and tasks, with substantial headroom for future improvement.
The benchmark, dataset, and evaluation harness are publicly available at \url{https://github.com/uw-math-ai/quantum-ai}.
\end{abstract}


% \thispagestyle{firstpagefooter}

%% ============================================================
%%  SECTION I — INTRODUCTION
%%  Owner: Andres | Editor: Sarju
%% ============================================================
\section{Introduction}
\label{sec:intro}

Quantum computing promises transformative speedups in domains such as cryptography~\cite{shor1997} and quantum chemistry simulation~\cite{aspuru2005}, yet writing quantum programs remains challenging.
Unlike classical computation, where programmers can build on decades of established abstractions, quantum programming demands reasoning about unintuitive phenomena.
Programs operate on qubits whose states are described by probability amplitudes with complex phases, and outcomes are inherently probabilistic.
The no-cloning theorem forbids copying quantum states, and measurement collapses superpositions, severely limiting what developers can observe during computation.
Although recent work has proposed statistical assertions~\cite{huang2019assertions} and runtime monitoring~\cite{ma2025qmon} for quantum programs, debugging remains far less mature than in classical programs, and recent studies find that over 80\% of quantum bugs are domain-specific~\cite{luo2022bugs, ramalho2024debugging}.

The programming languages available today provide limited abstraction over the underlying hardware.
Languages such as OpenQASM~\cite{cross2022openqasm3}, Q\#~\cite{svore2018qsharp}, and Guppy~\cite{guppy2024} express quantum computation as sequences of unitary operations applied to individual qubits, differentiating themselves primarily through their classical control structures rather than offering meaningful quantum abstractions.
More recent research languages like Qunity~\cite{voichick2023qunity} and Tower~\cite{yuan2024tower} have explored higher-level abstractions, but have not yet achieved the kind of intuitive programming model that would make quantum software development broadly accessible.
Meanwhile, optimizing compilers and specialized tools such as Stim~\cite{gidney2021stim} have advanced circuit-level automation, yet the burden of correct-by-construction design still falls largely on the programmer.

Large language model (LLM) agents offer a compelling alternative.
These systems have demonstrated remarkable capabilities in generating complex classical software, from solving competitive programming problems to autonomously resolving real-world software engineering tasks~\cite{jimenez2024swebench, chen2021humaneval}.
A natural aspiration is that agents could similarly be used to write quantum software: a user would describe a computational problem in natural language, and the agent would automatically identify which components admit quantum speedups, synthesize correct quantum code for those components, and integrate them into a hybrid classical-quantum solution---largely bypassing the abstraction gap.
To track progress toward this vision, the community needs rigorous benchmarks.

 Several quantum code-generation benchmarks have recently emerged~\cite{qiskit_humaneval, quanbench, qhackbench, qcoder}, but none target stabilizer-circuit synthesis, and most rely on exponentially costly state-vector verification that limits scalability.
Moreover, no existing benchmark addresses the specialized and practically critical domain of \emph{quantum error correction (QEC) circuit synthesis}, including stabilizer codes, circuit optimization under semantic constraints, and fault-tolerant circuit generation, which demands domain-specific metrics such as fault-tolerance scores and error propagation analysis.

In this work, we introduce \textsc{StabilizerBench}, a benchmark for AI-assisted quantum circuit synthesis grounded in stabilizer circuits.
We focus on stabilizer circuits for two reasons.
First, they are practically critical: stabilizer codes underpin quantum error correction, which is widely considered essential for achieving utility-scale fault-tolerant quantum computation~\cite{knill1997theory, terhal2015quantum}.
Second, they serve as an excellent proxy for general quantum circuit synthesis: constructing stabilizer circuits exercises the same core competencies required for arbitrary quantum programming, including gate decomposition, qubit routing, and semantic-preserving transformations, while admitting efficient polynomial-time verification~\cite{gottesman1998heisenberg, aaronson2004improved}.
Correctness can be checked by confirming that the output state is a $+1$~eigenstate of every stabilizer generator, without requiring exponentially costly simulation.
This property makes stabilizer circuits uniquely suited for a scalable benchmark.

\textsc{StabilizerBench} comprises 192 stabilizer codes spanning 14~families, 4--196~qubits, and distances 2--21, organized into three benchmark tasks of increasing difficulty:

\begin{itemize}
    \item \textbf{B1: State-Preparation Circuit Generation.}
    Given a set of stabilizers, produce a valid state-preparation circuit.
    This tests basic quantum programming competency---well-known algorithms exist for constructing such circuits~\cite{aaronson2004improved, cleve1997efficient}.

    \item \textbf{B2: Circuit Optimization.}
    Given a correct circuit and its stabilizers, produce a functionally equivalent circuit with fewer entangling gates and shallower depth.
    This tests structured reasoning about circuit equivalences and trade-offs across a vast design space.

    \item \textbf{B3: Fault-Tolerant Circuit Generation.}
    Given a non-fault-tolerant circuit, restructure it and add flag qubits so that single faults do not propagate into uncorrectable errors.
    Fault tolerance admits multiple slightly different definitions in literature, and synthesizing fault-tolerant state-preparation circuits remains a computationally hard problem even for known codes~\cite{Peham_2025}.
    This tests reasoning about error propagation and modifying circuit structure by adding new qubits without changing the semantics of the original circuit.
\end{itemize}

Following the design philosophy of classical benchmarks such as SWE-bench~\cite{jimenez2024swebench}, \textsc{StabilizerBench} specifies inputs, verification oracles, and scoring, but leaves prompts and agent strategies open.
This separation ensures the benchmark measures the full agent pipeline---model, prompt, and tool use---and remains durable as prompting techniques evolve.

We make the following contributions:
\begin{enumerate}
    \item \textbf{StabilizerBench}, a benchmark suite of 192 stabilizer codes with three tasks (B1--B3) and automated, polynomial-time verification oracles---the first benchmark targeting AI-assisted QEC circuit synthesis.

    \item \textbf{A unified, generator-weighted scoring system} with two tiers: a \emph{capability score} measuring breadth of success across the code suite, and a \emph{quality score} capturing circuit merit, both  weighted naturally by code complexity determined by the stabilizers.

    \item \textbf{A continuous fault-tolerance metric} that quantifies how close a circuit is to fault-tolerance, enabling graded comparison rather than binary pass--fail classification.

    \item \textbf{Baseline evaluation of three frontier AI agents}---Claude Opus~4.6, GPT-5.2, and Gemini~3 Pro Preview---validating that the benchmark is discriminative across models and tasks, with substantial headroom for future improvement.
\end{enumerate}

The remainder of this paper is organized as follows:
Section~\ref{sec:background} provides background on stabilizer codes, fault tolerance, and AI code generation benchmarks;
Section~\ref{sec:methodology} describes the design of \textsc{StabilizerBench}: its principles, code suite, tasks, verification oracles, scoring, and evaluation harness;
Section~\ref{sec:results} presents baseline evaluation results;
Section~\ref{sec:discussion} discusses implications, limitations, and future work, and Section~\ref{sec:conclusion} concludes.

%% ============================================================
%%  SECTION II — BACKGROUND & RELATED WORK (~1 page)
%% ============================================================
\section{Background and Related Work}
\label{sec:background}

%% ============================================================
%%  II-A. Stabilizer Formalism & QEC (0.3 pages)
%%  Writer: Cordelia | Editor: Christian
%% ============================================================
\subsection{Stabilizer Formalism and Quantum Error Correction}
\label{sec:bg-stabilizers}

% Cordelia: write this subsection.
% Content guidance (from project plan):
% - Pauli group, stabilizer generators, code parameters [[n,k,d]]
% - State preparation circuits: map |0>^n to the codespace
% - Why stabilizer circuits: efficiently simulable (Gottesman-Knill), easy to verify
%
% NOTE: The old version is preserved in sections.old/background.tex.
% You can reuse the stabilizer formalism content from there.
We briefly introduce basic notions from quantum computing and error correction.  For a more detailed description, see \cite{nielsen2010quantum}.

Quantum computation uses \emph{qubits}, with basis states
\(
\ket{0}=\begin{pmatrix}1 & 0\end{pmatrix}^T\) and 
\(\ket{1}=\begin{pmatrix}0 & 1\end{pmatrix}^T.
\)
A qubit state is a superposition
\(
\ket{q}=\alpha\ket{0}+\beta\ket{1}\) with 
\(
\alpha,\beta \in \mathbb{C}
\) and
\(
|\alpha|^2+|\beta|^2=1.
\)
Multi-qubit systems are described by tensor products, and quantum gates are unitary operators.
An important family of operators is the \emph{Pauli group}, generated by Pauli matrices $I, X, Y, Z$.
The Pauli operators either commute or anticommute, and quantum noise is often modeled as unintended Pauli errors that occur during the computation.
The \emph{\(n\)-qubit Pauli group} \(\mathcal{P}_n\) is the group of all \(n\)-fold tensor products of \(I, X, Y, Z\), together with the global phases \(\pm 1\) and \(\pm i\). An \(n\)-qubit unitary \(U\) is called a \emph{Clifford gate} if for every \(P \in \mathcal{P}_n\), we have \(UPU^\dagger \in \mathcal{P}_n\)
up to an overall phase.

% Let \(S \subseteq \mathcal{P}_n\) be an abelian subgroup with \(-I \notin S\), generated by
% \(
% S=\langle g_1,\dots,g_r\rangle.
% \)
% The subgroup \(S\) is called a \emph{stabilizer group}, and the elements \(g_1,\dots,g_r\) are called \emph{stabilizer generators}, or just \emph{stabilizers}. The associated \emph{stabilizer code} is the code space
% \[
% \mathcal{C}(S)=\left\{\ket{\psi}: g_i\ket{\psi}=\ket{\psi}\text{ for all } 1\le i\le r\right\}.
% \]
An abelian subgroup \(S\subseteq \mathcal{P}_n\) with \(-I\notin S\) is called a \emph{stabilizer group}. If 
\(
S=\langle g_1,\dots,g_r\rangle,
\)
then \(g_1,\dots,g_r\) are called \emph{stabilizer generators}, or simply \emph{stabilizers}. The associated \emph{stabilizer code} is the codespace
\[
\mathcal{C}(S)=\left\{\ket{\psi}: g_i\ket{\psi}=\ket{\psi}\text{ for all } 1\le i\le r\right\}.
\]
Equivalently, the codespace is the joint \(+1\)-eigenspace of the stabilizers.

A stabilizer code is called an \emph{\([[n,k,d]]\)-code} if it is a subspace of \((\mathbb{C}^2)^{\otimes n}\) encoding \(k\) \emph{logical qubits} into \(n\) \emph{physical qubits}, with \emph{code distance} \(d\). The code space has dimension \(2^k\), and \(d\) is the minimum weight of a Pauli operator that maps a codeword to another indistinguishable logical state.
A \emph{state preparation circuit} for a stabilizer code maps the initial state \(\ket{0}^{\otimes n}\) to a state in the codespace. More generally, such a circuit prepares encoded logical states from the standard computational-basis input.

% We focus on stabilizer codes as they are both computationally practical and mathematically easy to verify; see \cite{Gottesman1997} and \cite{AaronsonGottesman2004}. By the Gottesman-Knill theorem, their action can be simulated efficiently on a classical computer, and their correctness can be checked using the stabilizer formalism; see \cite{NielsenChuang2010} for details.

We focus on stabilizer codes because they are both computationally practical and mathematically straightforward to verify \cite{Gottesman1997, aaronson2004improved}. By the Gottesman--Knill theorem \cite{gottesman1998heisenberg}, any $n$-qubit stabilizer circuit can be classically simulated in $O(n^2)$ time by tracking its tableau representation: a compact data structure that records how each Pauli generator transforms under the circuit's gates. Correctness of a state-preparation circuit can then be checked by confirming that the output tableau stabilizes every generator of the target code, without requiring exponentially costly state-vector simulation.
%% ============================================================
%%  II-B. Fault Tolerance (0.3 pages)
%%  Writer: Christian | Editor: Cordelia
%% ============================================================
\subsection{Fault Tolerance}
\label{sec:bg-ft}

% Christian: write this subsection.
% Content guidance (from project plan):
% - Definition: single faults must not cause uncorrectable errors
% - Flag qubits and flagged circuits
% - The gap: no universally adopted continuous FT metric
%
% NOTE: The old version is preserved in sections.old/background.tex.
% You can reuse the fault tolerance definitions from there.
While quantum error correcting codes enable a significant reduction in erroneous quantum computations, it is non-trivial to prepare a logical encoding of the qubits on noisy hardware in a \emph{fault tolerant} way.
Fault tolerance does not have a standardized definition in literature, so we consider fault tolerance as the ability to detect whether a catastrophic fault has occurred, allowing circuit execution to be aborted and then restarted if needed.

More formally, let \(C\) be a stabilizer circuit on $n$ qubits that takes \(\ket{0}^{\otimes n}\) as input.
% \begin{definition}[Fault at location $i$]
%     A Pauli $F(i,P)\in \mathcal{P}_n$ such that $$F(i,P) = \left(\bigotimes_{k=1}^{i-1} I\right) \otimes P \otimes \left(\bigotimes_{k=i+1}^{n} I\right)$$
%     where $i\in \{1,2,...,n\}$ and $P\in \{X,Y,Z\}$ is a \emph{fault}.
% \end{definition}

\begin{definition}[Fault]
Let $i\in \{1,2,\dots,n\}$ and $P\in \{X,Y,Z\}$. A fault on qubit $i$, denoted by $F_{i,P}$ is the $n$-qubit Pauli operator acting as $P$ on qubit $i$ and as the identity on all other qubits, i.e.
\[F_{i,P}  = \left(\bigotimes_{k=1}^{i-1} I\right) \otimes P \otimes \left(\bigotimes_{k=i+1}^{n} I\right).\]
\end{definition}

Let $m$ be the depth of $C$ and let $L(C) = \{1,2,...,n\}\times \{0,1,...,m\}$.
Let
\[
C = C_{\mathrm{suf}}^{(j)} C_{\mathrm{pre}}^{(j)}
\]
be the decomposition of \(C\) at layer \(j\), where \(C_{\mathrm{pre}}^{(j)}\) consists of the first \(j\) layers and \(C_{\mathrm{suf}}^{(j)}\) consists of the remaining \(m-j\) layers.

We write
\[
\mathcal{S}(C)
=
\left\{
\,C_{\mathrm{suf}}^{(j)} F_{i,P} C_{\mathrm{pre}}^{(j)}
:\ (i,j)\in L(C),\ P\in\{X,Y,Z\}
\right\}
\]
for the set of all single-fault variants of \(C\).

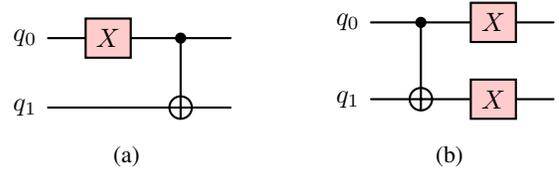
\begin{figure}
    \centering
    % Subfigure a
    \begin{subfigure}[b]{0.23\textwidth}
        \centering
        \begin{quantikz}
        \lstick{$q_0$} & \gate[style={fill=red!20}]{X}
        &  \ctrl{1} & \\
        \lstick{$q_1$} &   & \targ{} &
        \end{quantikz}  
        \label{fig:single-fault}
        \caption{}
    \end{subfigure}
    % Subfigure b
    \begin{subfigure}[b]{0.23\textwidth}
        \centering
        \begin{quantikz}
        \lstick{$q_0$} 
        &  \ctrl{1} & \gate[style={fill=red!20}]{X} & \\
        \lstick{$q_1$} &   \targ{} & \gate[style={fill=red!20}]{X} & 
        \end{quantikz}  
        \label{fig:two-error}
        \caption{}
    \end{subfigure}
    % \begin{quantikz}
    %     \lstick{$q_0$} & \gate[style={fill=red!20}]{X} & \ctrl{1} & \\
    %     \lstick{$q_1$} & & \targ{} &
    %     \end{quantikz}
    %     =\begin{quantikz}
    %     \lstick{$q_0$} & \ctrl{1} & \gate[style={fill=red!20}]{X} & \\
    %     \lstick{$q_1$} & \targ{} & \gate[style={fill=red!20}]{X} &
    % \end{quantikz}
    \caption{(a) A single \(X\) fault on $q_0$ occurring before the CNOT gate can be equivalently expressed in (b) as an \(X \otimes X\) error on the two qubits after the CNOT. We say (b) is the result of \emph{propagating} the fault in (a).} 
    \label{fig:X_fault_prop}
\end{figure}

\begin{definition}[Error]
Let \(C' \in \mathcal{S}(C)\). The error associated with \(C'\) is the operator
\[
E := C' C^\dagger,
\]
or equivalently, $C'=EC$. We refer to the process of determining \(E\) from the inserted fault $F_{i,P}$ as fault propagation.
\end{definition}
Note that since $C_{\mathrm{suf}}^{(j)}, C_{\mathrm{pre}}^{(j)}$ are both unitary, we have
\[
E = C' C^\dagger
  = \bigl(C_{\mathrm{suf}}^{(j)} F_{i,P} C_{\mathrm{pre}}^{(j)}\bigr)
    \bigl(C_{\mathrm{suf}}^{(j)} C_{\mathrm{pre}}^{(j)}\bigr)^\dagger
  = C_{\mathrm{suf}}^{(j)} F C_{\mathrm{suf}}^{(j)\dagger}.
\]
Since $C_{\mathrm{suf}}^{(j)} F C_{\mathrm{suf}}^{(j)\dagger}$ is a conjugation by a Clifford circuit, we have \(E \in \mathcal{P}_n\). We give an example of fault propagation in Figure~\ref{fig:X_fault_prop}. 

% Because \(F \in \mathcal{P}_n\) and conjugation by a Clifford circuit preserves the Pauli group, we conclude that \(E \in \mathcal{P}_n\).

% Note that for any $E\in \mathcal{P}_n$, we have
% \[
%     E = C'C^{\dagger} = (C_{suf}FC_{pre})(C_{suf}C_{pre})^{\dagger} = C_{suf}FC_{suf}^{\dagger}
% \]
% Since $F\in\mathcal{P}_n$, by Clifford conjugation, $E\in\mathcal{P}_n$.

\begin{definition} [Accept]
    Let \[\pi: \{C\} \cup \mathcal{S}(C) \rightarrow \{0,1\}\] be a binary decision rule.
    We accept a circuit \mbox{$C'\in \{C\}\cup \mathcal{S}(C)$} if $\pi(C') = 0$.
    We reject $C'$ if $\pi(C') = 1$.
\end{definition}

Let \(W(C,C')\) denote the weight of the error \mbox{$E = \bigotimes_{i=1}^n P_i$} for $P_i\in \mathcal{P}$, i.e.,
$$W(C,C')=\left|\{i\in\{1,2,...,n\}:P_i\neq I\}\right|$$
\begin{definition} [Fault Tolerance]
    Let \(t\ge 0\) be an error threshold and $\pi$ be a decision rule.
    We say that a circuit \(C\) is fault tolerant up to threshold \(t\) with respect to $\pi$ if $\pi(C) = 0$ and
    \[
    \max_{C'\in \mathcal{S}(C)}
    \Bigl(W(C,C')\cdot (1-\pi(C'))\Bigr)\le t.
    \]
\end{definition}
For a stabilizer code of distance $d$, we take the error threshold to be 
$t = \left\lfloor \frac{d-1}{2} \right\rfloor$,
since such a code can uniquely correct up to 
$\left\lfloor \frac{d-1}{2} \right\rfloor$ errors~\cite{Gottesman1997}.

% The main fault-tolerance strategy we emphasize is the use of \emph{flag qubits}.
% Using various flag techniques \cite{Prabhu_2023, Peham_2025}, we can entangle the data qubits with the flag qubits in such a way that if errors propagate between data qubits, they can also propagate to flag qubits and subsequently be detected.
% If we measure our flag qubits in the computational basis, and each measurement outputs "$0$", then we conclude that no uncorrectable errors have occurred, and \emph{accept} the stabilizer state. Otherwise, we \emph{reject} the state and re-execute the preparation circuit.
% See Figure \ref{fig:flag} for an example of flagging.

\begin{figure}
    % \centering

    % Subfigure a
    \begin{subfigure}[b]{0.23\textwidth}
        \centering
        \scalebox{0.73}{\begin{quantikz}
            \lstick{$q_0$} & \ctrl{2} & \gate[style={fill=red!20}]{X} & \ctrl{1} & \ctrl{2} & \\
            \lstick{$q_1$} & & & \targ{} &  &  \\
            \lstick{$f = \ket{0}$} & \targ{} & & & \targ{}  &\meter{} \\
        \end{quantikz}}
        \label{fig:single-fault-flag}
        \caption{}
    \end{subfigure}
    \quad
    % Subfigure b
    \begin{subfigure}[b]{0.23\textwidth}
        \centering
        \scalebox{0.73}{\begin{quantikz}
            \lstick{$q_0$} & \ctrl{2} & \ctrl{1} & \ctrl{2} & \gate[style={fill=red!20}]{X} & \\
            \lstick{$q_1$} & & \targ{} &  & \gate[style={fill=red!20}]{X} & \\
            \lstick{$f = \ket{0}$} & \targ{} & & \targ{} & \gate[style={fill=red!20}]{X} &\meter{} \\
        \end{quantikz}}
        \label{fig:three-error-flag}
        \caption{}
    \end{subfigure}

    \caption{The pair of CNOT gates from $q_0$ to $f$, which form this \emph{flag gadget} act as a bit-parity check for $q_0$. (a) An $X$ fault that occurs on the $q_0$ wire in between this pair of CNOT gates propagates as an $X$ error on $f$ as shown in (b), which can be measured in the $Z$ basis, outputting a 1, which means that $q_0$ has an odd bit-parity with itself across the flag gadget where even parity is expected, indicating the presence of an $X$ fault.}
    \label{fig:flag}
\end{figure}
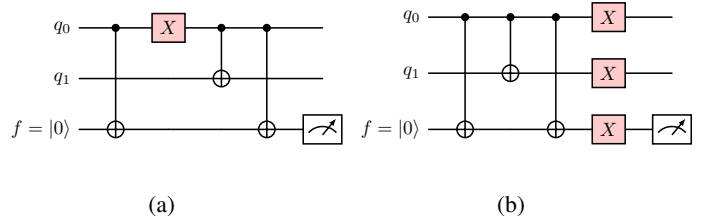

Using flag-based techniques \cite{PhysRevLett.121.050502, Chamberland2018flagfaulttolerant, Prabhu_2023, Peham_2025}, we can entangle the data qubits with auxiliary flag qubits so that error propagation to the data can trigger a detectable change on the flag qubits as well. We then measure the flag qubits in the computational basis and use the decision rule \(\pi(C')=1\) if at least one flag qubit is measured to be \(1\), and \(\pi(C')=0\) otherwise. See Figure~\ref{fig:flag} for an example of flagging.

% \begin{figure}
%     \centering
%     \scalebox{0.75}{
%     \begin{quantikz}
%         \lstick{$q_0$} & \ctrl{2} & \gate[style={fill=red!20}]{X} & \ctrl{1} & \ctrl{2} & \\
%         \lstick{$q_1$} & & & \targ{} &  &  \\
%         \lstick{$f$} & \targ{} & & & \targ{}  &\meter{} \\
%     \end{quantikz}=
%     \begin{quantikz}
%         \lstick{$q_0$} & \ctrl{2} & \ctrl{1} & \ctrl{2} & \gate[style={fill=red!20}]{X} & \\
%         \lstick{$q_1$} & & \targ{} &  & \gate[style={fill=red!20}]{X} & \\
%         \lstick{$f$} & \targ{} & & \targ{} & \gate[style={fill=red!20}]{X} &\meter{} \\
%     \end{quantikz}
%     }
%     \caption{If an $X$ fault propagates through the flag gadget, the fault can be detected by measuring the flag qubit.}
%     \label{fig:flag}
% \end{figure}
%% ============================================================
%%  II-C. Benchmarking AI Code Generation (0.2 pages)
%%  Writer: Sarju | Editor: Andres
%% ============================================================
\subsection{Benchmarking AI Code Generation}
\label{sec:bg-benchmarks}

Automated evaluation of AI-generated code began with functional correctness 
benchmarks for classical programs. HumanEval~\cite{chen2021humaneval} introduced 
hand-written Python problems with unit-test oracles; MBPP~\cite{austin2021mbpp} 
extended this to a larger corpus of crowd-sourced tasks. SWE-bench~\cite{jimenez2024swebench} 
raised the difficulty bar by grounding tasks in real GitHub issues, requiring 
agents to reason over full repository contexts. These benchmarks share a 
common design principle: verification is cheap and deterministic, enabling 
automated, scalable evaluation.

Quantum code generation benchmarks have followed a similar trajectory. 
Qiskit HumanEval~\cite{qiskit_humaneval} adapts the HumanEval format to Qiskit, testing 
whether agents can produce correct quantum circuit implementations of 
standard subroutines. QCoder~\cite{qcoder} and QuanBench~\cite{quanbench} broaden 
coverage to a wider range of quantum programming tasks, while 
QHackBench~\cite{qhackbench} draws from competition-style problems. However, 
these benchmarks share three limitations that make them ill-suited for 
evaluating QEC circuit synthesis. First, tasks are general-purpose — they 
test knowledge of quantum gates and standard algorithms, not the 
domain-specific reasoning required for stabilizer code manipulation. Second, 
verification relies on full state-vector simulation, whose cost grows 
exponentially with qubit count, limiting benchmarks to small circuits. 
Third, none of these benchmarks expose the semantic constraints central to 
QEC: that a valid circuit must preserve a specific stabilizer group, 
optimize under error-propagation metrics, and tolerate physical faults.

To our knowledge, no existing benchmark targets the synthesis, optimization, 
or fault-tolerant generation of stabilizer code circuits — the tasks most 
critical for near-term quantum error correction. \textsc{StabilizerBench} fills this gap.

% Sarju: write this subsection.
% Content guidance (from project plan):
% - Classical code benchmarks (HumanEval, SWE-bench, MBPP)
% - Quantum code-gen benchmarks: Qiskit HumanEval, QuanBench, QCoder, QHackBench
%   — limitations: general-purpose tasks, small scale, exponential-cost verification
% - Benchmark design principles (Rohe et al. 2025)
% - Gap: no benchmark for QEC circuit synthesis

%% ============================================================
%%  II-D. Tools & Frameworks (0.2 pages)
%%  Writer: Christian | Editor: Mayee
%% ============================================================
\subsection{Tools and Frameworks}
\label{sec:bg-tools}

% Christian: write this subsection.
% Content guidance (from project plan):
% - Stim simulator and stabilizer tableau propagation
% - GitHub Copilot SDK and CLI
%
% NOTE: The old version is preserved in sections.old/background.tex.
% You can reuse the Stim and Copilot SDK content from there.
\subsubsection{Stim}
The advantage of stabilizer QECC is that stabilizer circuits can be simulated efficiently classically \cite{aaronson2004improved}.
We use the Stim \cite{gidney2021stim} library to construct, simulate, and verify state preparation circuits for a wide range of stabilizer codes [\ref{sec:code-suite}] on up to 196 qubits, which is computationally intractable for general-purpose quantum simulation libraries such as Qiskit.
We require efficient simulation of our quantum circuits in order to quickly verify that an agent's output correctly prepares the given stabilizer state and is fault tolerant.
Stim also provides a clean, human-readable and agent-friendly text representation of the quantum circuits, which simplifies translation between circuit objects in code and LLM inputs and outputs.
\subsubsection{GitHub Copilot SDK}
We integrate the GitHub Copilot SDK into our codebase in order to:
\begin{itemize}
    \item Prompt LLM agents via code for repeatable and automated testing
    \item Create a feedback loop for agents to verify their outputs via tools that incorporate Stim
    \item Leverage the diverse suite of LLM agents, such as GPT, Claude, and Gemini
    \item Enable agents to make local changes, such as creating temporary files, if they required such resources
\end{itemize}

%% ============================================================
%%  SECTION III — STABILIZERBENCH: BENCHMARK DESIGN (~1.75 pages)
%% ============================================================
\section{StabilizerBench: Benchmark Design}
\label{sec:methodology}

%% ============================================================
%%  III-A. Design Principles (0.25 pages)
%%  Writer: Cordelia | Editor: Andres
%% ============================================================
\subsection{Design Principles}
\label{sec:design-principles}

% Cordelia: write this subsection.
% Content guidance (from project plan):
% - Map six benchmark quality criteria (Rohe et al. 2025) to our design choices:
%   Relevance (QEC is critical), Fairness (model-agnostic harness),
%   Reproducibility (deterministic oracles), Usability (plug-your-model-in),
%   Scalability (4–196 qubits), Transparency (mathematical scoring)

% Following the benchmark quality criteria summarized by Rohe et al.~\cite{rohe2025quantum}, our benchmark is designed to satisfy six key properties. It is \emph{relevant} because quantum error correction is a central challenge in quantum computing. Our benchmark includes circuits ranging from 4 to 196 qubits, and we are able to also add new circuits in the future and scale up the benchmark size without affecting the existing score.  
% It is \emph{reproducible} because our benchmark provides people with the setup, constraints, and code to run benchmark themselves. 
% It is \emph{fair} because all models are evaluated under the same model-agnostic harness.
% It is \emph{verifiable} because one can conduct circuit validation and error analysis to verify all result.
% It is \emph{usable} because the framework is designed so that one can readily plug in a new model. 

Following the benchmark quality criteria summarized by Rohe et al.~\cite{rohe2025quantum}, our benchmark is designed to satisfy the following five key properties:
\begin{enumerate}
    \item \emph{Relevance}: Quantum error correction is a central challenge in quantum computing, and our benchmark evaluates model performance on this task across circuits ranging from 4 to 196 qubits.
    
    \item \emph{Reproducibility}: We provide the complete setup, constraints, prompts, and codes required to rerun the benchmark under the same test configuration and reproduce the results.
    
    \item \emph{Fairness}: All models are evaluated within the same model-agnostic harness.
    
    \item \emph{Verifiability}: Outputs can be checked through circuit validation and error analysis using transparent and well-defined procedures.
    
    \item \emph{Usability}: The framework is straightforward to run and readily accommodates new models without substantial modification.
\end{enumerate}
%% ============================================================
%%  III-B. Code Suite (0.25 pages)
%%  Writer: Sarju | Editor: Cordelia
%% ============================================================
\subsection{Code Suite}
\label{sec:code-suite}

% Sarju: write this subsection.
% Content guidance (from project plan):
% - 192 stabilizer codes: 24 base codes + 168 tensor products
% - Code families, parameter ranges, coverage analysis
% - Distribution across qubit counts, distances, families

%% ============================================================
%%  III-B. Code Suite (0.25 pages)
%%  Writer: Sarju | Editor: Cordelia
%% ============================================================

StabilizerBench is built on a dataset of 192 stabilizer codes drawn from
diverse families and scaled to cover a wide range of circuit complexities.

\emph{Base codes.}
The dataset includes 24 base codes spanning six families:
\emph{rotated surface codes} ($d \in \{3,5,7\}$)~\cite{tomita2014rotated},
\emph{color codes} (hexagonal~\cite{bombin2006topological} and
square-octagon~\cite{landahl2011fault}, $d \in \{3,5,7\}$),
\emph{Iceberg codes} ($m \in \{2,3,4\}$)~\cite{self2024iceberg},
\emph{many-hypercube codes} ($\ell \in \{1,2\}$)~\cite{goto2024hypercube},
\emph{bivariate bicycle (BB) codes}
($n \in \{72, 90\}$)~\cite{bravyi2024high},
and a set of well-established codes including the
Perfect 5-Qubit~\cite{laflamme1996perfect},
Steane, Hamming, and Golay~\cite{steane1996multiple},
Shor~\cite{shor1995scheme},
Tetrahedral~\cite{steane1996reed_muller},
Carbon~\cite{knill2005realistic,paetznick2024logical},
and 4-Qubit Detector~\cite{Gottesman1997} codes.

\emph{Tensor products.}
The remaining 168 circuits are tensor products of pairs of base codes,
constructed by taking the joint stabilizer group of two independent
subsystems.  Because stabilizer groups on disjoint qubit supports
compose as a direct product~\cite{Gottesman1997,nielsen2010quantum},
tensor products preserve the stabilizer structure of each component
while systematically increasing circuit complexity without requiring
new code designs. This construction also introduces cross-family combinations (e.g., surface $\otimes$ color, Shor $\otimes$ Golay) that stress-test
generalization beyond any single code family.

\emph{Coverage.}
Across all 192 circuits, stabilizer counts range from 2 to 194 with a
median of 80, providing a smooth difficulty gradient from trivial
instances to large codes that challenge frontier models.
The total number of stabilizer generators across the full suite is
$K = \sum_{i=1}^{192} k_i = 16{,}340$; this serves as the maximum
achievable score for all benchmarks.
Table~\ref{tab:code-suite} summarizes the base code families and their
stabilizer ranges.

% \begin{table}[t]
% \centering
% \caption{Base code families in StabilizerBench (24 codes).}
% \label{tab:code-suite}
% \begin{tabular}{lcc}
% \toprule
% Family & \# Codes & Stabilizer range \\
% \midrule
% Rotated Surface  & 3  & 8--48  \\
% Color (Hex + Sq-Oct) & 6 & 6--36 \\
% Iceberg          & 3  & 2--2   \\
% Hypercube        & 2  & 2--20  \\
% BB (LDPC)        & 2  & 72--90 \\
% Other (Steane, Shor, Golay, \ldots) & 8 & 2--22 \\
% \midrule
% \textbf{Total base} & \textbf{24} & \textbf{2--90} \\
% Tensor products  & 168 & 18--194 \\
% \midrule
% \textbf{Total} & \textbf{192} & \textbf{2--194} \\
% \bottomrule
% \end{tabular}
% \end{table}

\begin{table}[t]
\centering
\caption{Base code families in StabilizerBench (24 codes).}
\label{tab:code-suite}
\resizebox{\columnwidth}{!}{%
\begin{tabular}{lcc}
\toprule
Family & \# Codes & Stabilizer range \\
\midrule
Rotated Surface              & 3  & 8--48   \\
Color (Hex + Sq-Oct)         & 6  & 6--36   \\
Iceberg                      & 3  & 2--2    \\
Hypercube                    & 2  & 2--20   \\
BB (LDPC)                    & 2  & 72--90  \\
Other (Steane, Shor, Golay, \ldots) & 8  & 2--22   \\
\midrule
\textbf{Total base}          & \textbf{24}  & \textbf{2--90}   \\
Tensor products              & 168 & 18--194 \\
\midrule
\textbf{Total}               & \textbf{192} & \textbf{2--194}  \\
\bottomrule
\end{tabular}}
\end{table}

%% ============================================================
%%  III-E. Verification Oracles (0.25 pages)
%% ============================================================
\subsection{Verification Oracles}
\label{sec:oracles}

\textsc{StabilizerBench} provides polynomial-time
verification oracles, each exposed as an agent tool in an
automated feedback loop.

\subsubsection{Verifying Stabilizer Preservation}
Given a candidate circuit~$C$ and a stabilizer generator set
$\{s_1,\ldots,s_k\}$, this oracle runs Stim's tableau simulator on $C$
and checks whether the output state is a $+1$ eigenstate of each $s_j$.
If every generator is preserved, the full stabilizer group is preserved
by composition.
\emph{Returns:} per-generator pass/fail list and an overall validity flag.

\subsubsection{Propagating Faults}
\label{subsec:fault-prop}
Given a circuit~$C$ and a fault location, this oracle computes the
propagated Pauli error by applying the suffix circuit (everything after
the fault injection point) as a Clifford transformation to the injected
single-qubit Pauli~(see Section~\ref{sec:bg-ft}).
\emph{Returns:} the data-qubit error weight and the flag-qubit measurement
outcomes.

% Consider two qubits, $q_0, q_1$.
% Suppose a Pauli $X$ fault occurs on $q_0$ and the following suffix circuit is a CNOT with $q_0$ as the control and $q_1$ as the target.
% An equivalent circuit is first the same CNOT is applied, followed by $X$ errors on both $q_0$ and $q_1$.
% \begin{figure}[h]
%     \centering

%     % Subfigure a
%     \begin{subfigure}[b]{0.23\textwidth}
%         \centering
%         \includegraphics[width=\linewidth]{single-fault circuit.png}
%         \label{fig:single-fault}
%         \caption{}
%     \end{subfigure}
%     \hfill
%     % Subfigure b
%     \begin{subfigure}[b]{0.23\textwidth}
%         \centering
%         \includegraphics[width=\linewidth]{two-error circuit.png}
%         \label{fig:two-error}
%         \caption{}
%     \end{subfigure}

%     \caption{(a) A single $X$ fault on $q_0$ is equivalent to (b) an $X$ error on each of $q_0$ and $q_1$}
%     \label{fig:prop}
% \end{figure}
%% ============================================================
%%  III-C. Scoring Framework (0.25 pages)
%% ============================================================
\subsection{Scoring Framework}
\label{sec:scoring}

Every benchmark task $b \in \{1,2,3\}$ in \textsc{StabilizerBench} yields a score built from
up to two components: a \emph{capability score} ($S_{\mathrm{cap}}^{(b)}$) that measures whether the agent
can solve the task at all, and a \emph{quality score} ($S_{\mathrm{qual}}^{(b)}$) that additionally
measures how well it solves it.  Both are weighted by the number of
stabilizer generators~$k_i$ of code~$i$, so that harder codes with more
generators contribute proportionally more---no ad-hoc difficulty weights
are needed.

Let $N=192$ be the number of codes in the suite.  For each code~$i$, let
$k_i$ be its generator count and let $q_i\in[0,1]$ be a task-specific
quality factor (defined per benchmark below).  Then:
\begin{align}
    S_{\mathrm{cap}}^{(b)}  &= \sum_{i=1}^{N} \mathbf{1}[\text{agent succeeds on } i]\;\cdot\; k_i \label{eq:scap}\\[4pt]
    S_{\mathrm{qual}}^{(b)} &= \sum_{i=1}^{N} \mathbf{1}[\text{agent succeeds on } i]\;\cdot\; q_i \;\cdot\; k_i \label{eq:squal}
\end{align}
Because $q_i\in[0,1]$, we always have $S_{\mathrm{qual}}^{(b)} \leq S_{\mathrm{cap}}^{(b)}$.
The maximum achievable score for both metrics is
$K = \sum_{i=1}^{N} k_i = 16{,}340$, the total number of stabilizer
generators across the code suite.
Each benchmark defines its own notion of \emph{success} and its own
quality factor~$q_i$.

%% ============================================================
%%  III-D1. B1: State-Preparation Circuit Generation (0.15 pages)
%% ============================================================

\vspace{1em}
\subsubsection{B1: State-Preparation Circuit Generation}
\label{sec:task-b1}

B1 tests whether an agent can synthesize a quantum circuit that prepares a specified stabilizer state. Given only the stabilizer generators of a quantum error-correcting code, the agent must produce a Clifford circuit whose output is a $+1$ eigenstate of every generator—a task that requires navigating the combinatorial space of Clifford operations under global commutation constraints. While there exist classical algorithms that can programmatically generate such circuits, this benchmark evaluates whether an agent can independently derive or approximate these constructions without explicit procedural guidance.

\emph{Input:} The set of $k$ Pauli stabilizer generators
$\{s_1,\ldots,s_k\}$ for an $n$-qubit code.

\emph{Output:} A candidate Stim circuit~$C$ over $n$ qubits with initial
state $\lvert 0\rangle^{\otimes n}$.

\emph{Validity:} $C$ is valid if and only if its output state is a $+1$
eigenstate of every generator.  

% The verification oracle
% (Section~\ref{sec:oracles}) reconstructs the stabilizer tableau of $C$ via
% Gottesman--Knill simulation and checks each generator exactly.

\emph{Scoring:} B1 is a pure capability benchmark: $q_i = 1$ for every
successful instance.  The capability and quality scores coincide
(Equations~\ref{eq:scap}--\ref{eq:squal}).
To provide finer-grained diagnostics, we also report the number of
individually satisfied generators per code, even when the full set is not
achieved.

\emph{Agent interface:}
The agent is provided with a single agent tool,
\texttt{check\_stabilizers}, which it may invoke at most $A$ times.
\emph{Expects:} a Stim circuit string and the list of Pauli stabilizer
generators.
\emph{Returns:} a per-generator pass/fail list, the total number of
satisfied generators, and an overall validity flag.
Each invocation consumes one attempt.
%% ============================================================
%%  III-D2. B2: Circuit Optimization (0.15 pages)
%% ============================================================

\vspace{1em}
\subsubsection{B2: Circuit Optimization}
\label{sec:task-b2}

B2 tests whether an agent can reason about circuit equivalence to produce a
more efficient implementation of the same stabilizer state.  The agent is
given a \emph{baseline} Stim circuit and the list of $k$ stabilizer generators
it prepares, and must return a circuit that is both semantically equivalent
and strictly cheaper.

\emph{Input:} A valid Stim circuit $C_{\text{base}}$ and the set of
$k$ Pauli stabilizer generators $\{s_1,\ldots,s_k\}$.

\emph{Output:} A candidate Stim circuit $C'$ over the same qubit indices.

\emph{Validity:}  $C'$ is valid if and only if its stabilizer tableau
preserves all $k$ generators.  

\emph{Cost and strict improvement:}  Circuits are ordered by the
lexicographic cost tuple $(G_{2Q},\,D)$, where $G_{2Q}$ is the total
count of multi-qubit entangling gates (\textsc{CX}, \textsc{CZ}, \textsc{SWAP},
etc.) and $D$ is the circuit depth (minimum number of time steps under
maximum parallelism).
$G_{2Q}$ takes first priority because every entangling gate is an independent
noise channel; counting all multi-qubit gate types prevents the degenerate
strategy of substituting one entangling gate for another without reducing
entanglement overhead.  A valid candidate $C'$ is an \emph{improvement} if
and only if
\[
  \bigl(G_{2Q}(C'),\,D(C')\bigr)
    \;<\;
  \bigl(G_{2Q}(C_{\text{base}}),\,D(C_{\text{base}})\bigr)
\]
under lexicographic order.  Equality on both metrics is not an improvement.

\emph{Agent interface:}
The agent is provided with a single agent tool,
\texttt{evaluate\_\allowbreak optimization}, which it may invoke at
most $A$ times.
\emph{Expects:} a candidate Stim circuit string, the baseline circuit,
and the list of stabilizer generators.
\emph{Returns:} validity status, the number of preserved stabilizer
generators, the two-qubit gate count and depth of the candidate, and a
flag indicating whether the candidate is a strict improvement over the
baseline.
Each invocation consumes one attempt; no other tools are available
during optimization.

\emph{Scoring:}
The quality factor for B2 is the weighted optimization proportion:
\begin{align*}
  q_i &= 0.75\,\Delta^{G_{2Q}}_i + 0.25\,\Delta^{D}_i, \\
  \Delta^{m}_i &= \max\!\left(0,\,\min\!\left(1,\,
  \frac{b^m_i - o^m_i}{b^m_i}\right)\right)
\end{align*}
with $b^m_i$ and $o^m_i$ the baseline and optimized values of
metric $m \in \{G_{2Q},\, D\}$. The 3:1 weighting reflects the dominant noise contribution of entangling gates relative to depth. The capability and quality scores follow Equations~\ref{eq:scap}--\ref{eq:squal}.

%% ============================================================
%%  III-D3. B3: Fault-Tolerant Circuit Generation + FT Score (0.2 pages)
%% ============================================================
\vspace{1em}
\subsubsection{B3: Fault-Tolerant Circuit Generation}
\label{sec:task-b3}

B3 tests whether an agent can improve the fault tolerance of a given
circuit by inserting flag gadgets that detect uncorrectable error
propagation.  This task requires deep reasoning about how single-qubit
faults spread through entangling gates and how ancilla measurements can
catch dangerous fault paths.

\emph{Input:} A non-fault-tolerant Stim circuit~$C_{\text{base}}$, its
$k$ stabilizer generators $\{s_1,\ldots,s_k\}$, and the code
distance~$d$.

\emph{Output:} A candidate Stim circuit~$C'$ (possibly with additional
flag qubits) that preserves all $k$ generators and improves fault
tolerance.

\emph{Validity:} $C'$ is valid if and only if it preserves all $k$
stabilizer generators (checked identically to B1) and achieves a strictly
higher fault-tolerance score than $C_{\text{base}}$.

\emph{Fault-tolerance score.}
The fault tolerance (FT) score quantifies a circuit's ability to detect faults. It is defined on a normalized scale from 0 to 1, where 0 indicates no fault tolerance and 1 indicates full fault tolerance. 

Let $\mathcal{S}(C)$ denote the set of all single-fault variants of
circuit~$C$ (one single-qubit Pauli injected at each possible location).
For $C' \in \mathcal{S}(C)$, let $\pi\in\{0,1\}$ be the flag indicator:
\[\pi(C') = \begin{cases}
    0 & \text{if flags are up and circuit rejects,} \\
    1 & \text{if flags are down and circuit accepts,} \\
    1 & \text{if circuit has no flag.}
\end{cases}\]
Let $\mathcal{T}(\mathcal{S}(C))$ be the subset whose propagated error
weight exceeds the correctable threshold
$t = \lfloor(d{-}1)/2\rfloor$:
\[
\mathcal{T}(\mathcal{S}(C))
:=
\{\,C'\in \mathcal{S}(C)\mid W(C',C)>t\,\}.
\]
$\mathrm{FT}(C)\in[0,1]$ measures the proportion of dangerous fault
paths that are successfully flagged. If $\mathcal{T}(\mathcal{S}(C))=\emptyset$, set $\mathrm{FT}(C)=1$.
Otherwise:
\[
\mathrm{FT}(C)
=
\frac{1}{|\mathcal{T}(\mathcal{S}(C))|}
\sum_{C'\in \mathcal{T}(\mathcal{S}(C))}
\pi(C').
\]

\emph{Scoring:}
The quality factor for B3 is $q_i = \mathrm{FT}(C'_i)$, the
fault-tolerance score of the agent's best candidate circuit.
The capability and quality scores follow
Equations~\ref{eq:scap}--\ref{eq:squal}.

\emph{Agent interface:}
The agent is provided with a single agent tool,
\texttt{check\_fault\_tolerance}, which it may invoke at most $A$ times.
\emph{Expects:} a candidate Stim circuit string, the list of stabilizer
generators, and the code distance~$d$.
\emph{Returns:} validity status, the number of preserved stabilizer
generators, and the $\mathrm{FT}$ score of the candidate.
Each invocation consumes one attempt; the best-scoring valid candidate
is retained.
%% ============================================================
%%  III-F. Evaluation Harness (0.25 pages)
%% ============================================================
\subsection{Evaluation Harness}
\label{sec:harness}

% Christian: write this subsection.
% Content guidance (from project plan):
% - Architecture: User prompt -> CopilotClient -> LLM -> Tool calls -> Stim verification -> Feedback loop
% - MCP server exposing verification oracles as tools
% - The prompt is NOT part of the benchmark (SWE-bench model):
%   StabilizerBench defines inputs, oracles, and scoring — participants bring their own
%   prompts and agent strategies
% - Two configurable parameters: timeout (wall-clock limit per instance) and
%   validation attempts (iterations with oracle feedback). Results must report both.
% - Frame as "how to use the benchmark" — any agent can be plugged in, including local LLMs (Ollama)
% "Prompt can have significant impact on the result"
\begin{figure}
    \centering
    \includegraphics[width=0.8\columnwidth]{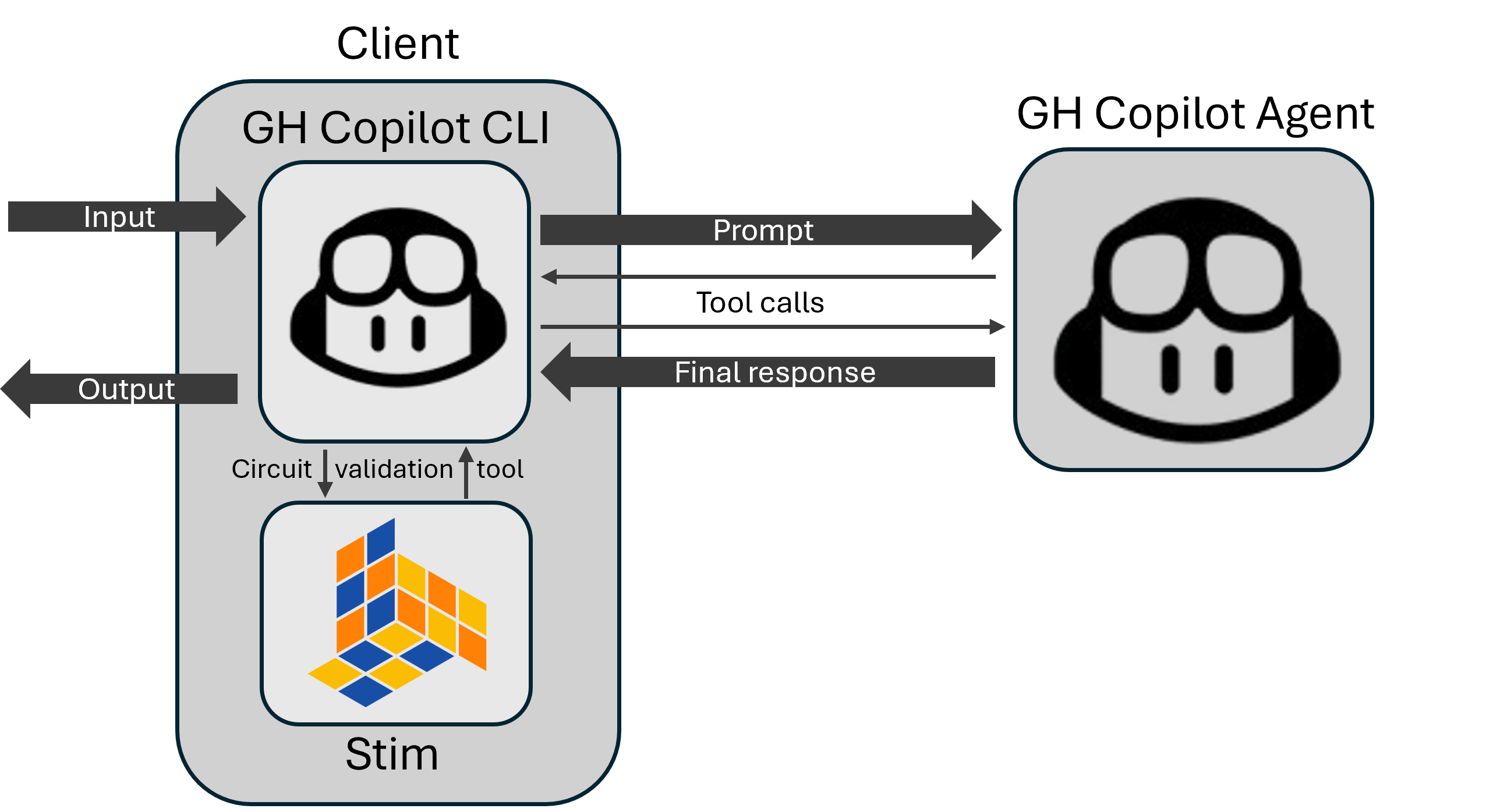}
    \caption{Agent workflow}
    \label{fig:agent}
\end{figure}
\textsc{StabilizerBench} provides a framework-agnostic evaluation harness that enables agents to leverage their thinking mode to iteratively refine quantum circuits via oracle feedback.
Our harness implements a feedback loop where agents can re-attempt generation using verification results to guide improvements.
\subsubsection{Architecture} The evaluation pipeline (see Fig. \ref{fig:agent}) follows an iterative workflow:
\begin{enumerate}
    \item[i)] the user selects a model and provides a task prompt [\ref{subsubsec:prompt}]
    \item[ii)] the harness connects the model to GitHub Copilot---either a model available out-of-the-box (e.g., GPT-5.2, Claude Opus 4.6) or a custom model connected through the Copilot extensibility layer
    \item[iii)] the agent generates or modifies a Stim circuit according to the task
    \item[iv)] the circuit is passed to verification oracles (Section~\ref{sec:oracles}) exposed as agent tools
    \item[v)] tool results are returned to the agent as structured feedback
    \item[vi)] the agent may iterate based on this feedback within a configurable attempt budget
\end{enumerate}
This workflow supports any model accessible through GitHub Copilot that can invoke tools and parse JSON responses.

\subsubsection{Configuration}
The user configures each benchmark run with:
\begin{itemize}
    \item \textbf{Model:} the LLM to evaluate, connected via GitHub Copilot.
    \item \textbf{Timeout:} a wall-clock time limit (in seconds) per code instance.
    \item \textbf{Attempt budget:} the maximum number of verification tool calls per instance. This defaults to 10.
\end{itemize}
Both the timeout and attempt budget will be reported alongside benchmark scores to enable fair comparison across evaluations.

\subsubsection{Prompt}
\label{subsubsec:prompt}
The harness provides a system prompt that handles tool invocation, attempt tracking, and output formatting; users do not need to instruct the agent to call verification tools.
The user-supplied task prompt needs only to include:
\begin{enumerate}
    \item[i)] task description and input parameters (stabilizers, baseline circuit, code distance, etc.)
    \item[ii)] the scoring objective for the benchmark
\end{enumerate}
Prompt design can have a significant impact on results and is deliberately left to the evaluator.

\subsubsection{Collecting Results}
The harness persists every benchmark run as a JSON file containing run-level metadata and per-instance results.
Run metadata records the model name, prompt, attempt budget, timeout, and wall-clock start/end timestamps.
For each code instance, the harness records every circuit the agent submits to a verification oracle along with the oracle's response---including the individual metrics like quality score, stabilizer preservation status, and per-stabilizer pass/fail details.

%% ============================================================
%%  SECTION IV — BASELINE EVALUATION (~2.5 pages)
%% ============================================================
\section{Baseline Evaluation}
\label{sec:results}

We evaluate three frontier models as baselines---Claude Opus 4.6 (Claude), GPT-5.2 (GPT), and Gemini 3 Pro Preview (Gemini)---on \textsc{StabilizerBench}. GPT-4.1 was only evaluated on B1. Due to its poor performance, further analysis on B2 and B3 is restricted to the three frontier models. 
\subsection{B1 Results: State-Preparation}
\label{sec:results-b1}

\begin{figure*}[t]
    \centering
    \includegraphics[width=\textwidth]{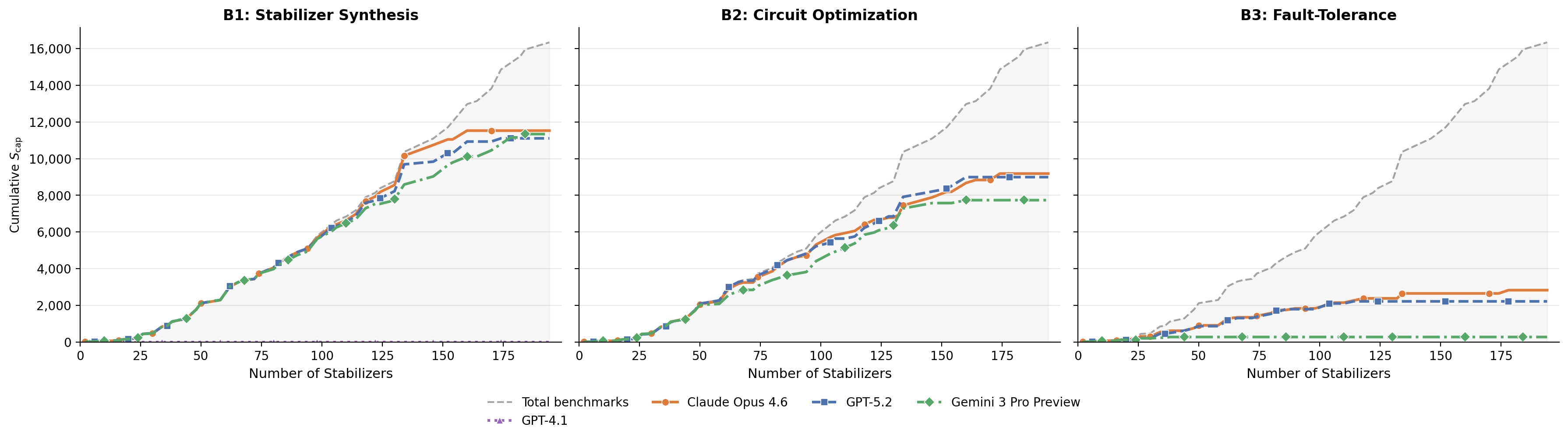}
    % \caption{Difficulty curves for all three tasks: cumulative performance vs.\ stabilizer count (best config per agent per code). \emph{B1}: codes perfectly synthesized; \emph{B2}: circuits successfully optimized (Claude and GPT at 15\,att/900\,s, Gemini at 1\,att); \emph{B3}: codes with positive FT score. All models degrade monotonically with circuit complexity; the gray dashed line shows the total benchmark ceiling.}
    \caption{Difficulty curves for all three tasks: cumulative capability score $S_\mathrm{cap}^{(b)}$ vs.\ stabilizer count under the best configuration for $b\in \{1,2,3\}$. All models degrade monotonically with circuit complexity; the gray dashed line shows the total benchmark ceiling.}
    \label{fig:difficulty-curves}
\end{figure*}

Our evaluation of three frontier AI agents on \textsc{StabilizerBench} shows strong capability in generating state-preparation circuits. Under the best configuration per code, all agents achieved high perfect-solve rates; with $S_{\mathrm{cap}}^{(1)}$ scores of 11{,}528, 11{,}106 and 11{,}340 (out of 16{,}340) respectively. Even when agents fail to fully solve a benchmark, they typically satisfy a substantial fraction of stabilizers, indicating that failures are structured near-misses rather than random outputs.

Configuration strongly affected performance, with timeout budget as the dominant factor. Increasing attempts from 1 to 15 under a 900 s timeout reduced aggregate $S_{\mathrm{cap}}^{(1)}$ from 30,776 to 24,294 out of 49,020 (3 × 16,340), with net $S_{\mathrm{cap}}^{(1)}$ losses across all agents ($-2,140$ Claude, $-2,404$ GPT, $-1,938$ Gemini), suggesting that iterative self-correction introduces overhead exceeding its benefit. In contrast, extending the 15-attempt timeout from 300\,s to 900\,s yielded 239 additional perfect solves and quadrupled the mean $S_{\mathrm{cap}}^{(1)}$ across all 3 agents from 2{,}021 to 8{,}098. The 300-second, 15-attempt setting performed worst overall. These results suggest that frontier models benefit more from sustained single-shot reasoning than from iterative refinement.

% \begin{table}[t]
% \centering
% \small
% \begin{tabularx}{\linewidth}{l l c c}
% \toprule
% \textbf{Model} & \textbf{Config} & \textbf{Perfect Solve Rate} & $\mathbf{S^{(1)}_{\text{cap}}}$ \\
% \midrule
% Claude & 1 att / 900s & 79.7\% & 10,772 \\
% Claude & 15 att / 300s & 25.0\% & 1,838 \\
% Claude  & 15 att / 900s & 72.9\% & 8,632 \\
% \midrule
% GPT & 1 att / 900s & 80.2\% & 10,646 \\
% GPT & 15 att / 300s & 24.0\% & 1,690 \\
% GPT & 15 att / 900s & 71.4\% & 8,242 \\
% \midrule
% Gemini  & 1 att / 900s & 72.4\% & 9,358 \\
% Gemini  & 15 att / 300s & 34.9\% & 2,534 \\
% Gemini  & 15 att / 900s & 64.1\% & 7,420 \\
% \bottomrule
% \end{tabularx}
% \caption{Performance comparison across models and configurations.}
% \label{tab:model_performance}
% \end{table}

\begin{table}[t]
\centering
\caption{B1 Circuit Synthesis Results. Summary of circuit capability scores across models and configurations. The configuration is listed
as attempts/timeout in seconds.}
\label{tab:model_performance}
\resizebox{\columnwidth}{!}{%
\begin{tabular}{llcc}
\toprule
Model & Config & Perfect Solve Rate & $S^{(1)}_{\mathrm{cap}}$ \\
\midrule
Claude & 1 att / 900s  & 79.7\% & 10{,}772 \\
Claude & 15 att / 300s & 25.0\% & 1{,}838 \\
Claude & 15 att / 900s & 72.9\% & 8{,}632 \\
\midrule
GPT    & 1 att / 900s  & 80.2\% & 10{,}646 \\
GPT    & 15 att / 300s & 24.0\% & 1{,}690 \\
GPT    & 15 att / 900s & 71.4\% & 8{,}242 \\
\midrule
Gemini & 1 att / 900s  & 72.4\% & 9{,}358 \\
Gemini & 15 att / 300s & 34.9\% & 2{,}534 \\
Gemini & 15 att / 900s & 64.1\% & 7{,}420 \\
\midrule 
GPT-4.1 & 15 att / 300s & 0.7 \% & 4 \\
\bottomrule
\end{tabular}}
\end{table}

We compared a detailed, guidance-heavy prompt (with algorithmic hints, budget-tracking rules, and strict tool-calling protocols) against a minimal prompt containing only the stabilizer generators, qubit count, and basic validation instructions. The minimal prompt consistently outperformed: GPT achieved a $S_{\mathrm{cap}}^{(1)}$ score of 4,526 vs. 1,690 out of 16,340, and Claude achieved 4,314 vs. 1,838. The gap widened on larger codes (16+ qubits), where the minimal prompt yielded $S_{\mathrm{cap}}^{(1)}$ scores of 4,264–4,468 compared to 1,646–1,788 for the detailed prompt. We attribute this to prescriptive hints anchoring models to strategies that fail to generalize, while added constraints consumed reasoning capacity and reduced exploration. This motivates \textsc{StabilizerBench}'s open-prompt design: embedding a specific prompt would measure prompt engineering rather than model capability.

Figure~\ref{fig:difficulty-curves} shows the B1 difficulty curves for all four models. While the other models solve a range of benchmark instances across increasing difficulty levels, GPT-4.1 succeeds on only a single case---the Perfect 5-Qubit Code (4 stabilizers)---and fails on all remaining tasks. This stark performance gap makes meaningful comparison difficult, so we exclude GPT-4.1 from the subsequent analysis.  The three frontier models exhibit steady gains up to roughly 160 stabilizers before plateauing: Claude and GPT-5.2 made no further gains beyond 174 and Gemini reaching its ceiling at 184. All agents achieve near-perfect success for codes with $\leq 100$ stabilizers (5{,}424 of 5{,}790 stabilizers solved by all three), but coverage drops sharply for 101--200 stabilizers (only 3{,}082 of 10{,}550 stabilizers solved by all three). Code distance shows a similar pattern, with near-universal success for $d \leq 10$ (6{,}224 of 7{,}268 stabilizers) but sharp degradation beyond (e.g., 302 of 1{,}914 stabilizers at $d = 14$). All 14 never-solved benchmarks are large tensor-product codes (148--196 qubits, $d \geq 12$), collectively accounting for 2{,}436 unsolved stabilizers.

%% ============================================================
%%  IV-B. B2 Results: Optimization (0.5 pages)
%%  Writer: Sarju | Editor: Sylvie
%% ============================================================
\subsection{B2 Results: Optimization}
\label{sec:results-b2}

Table~\ref{tab:b2} summarizes performance across all model--configuration
pairs on the 192-circuit optimization benchmark.

% \begin{table}[t]
% \centering
% \caption{B2 Circuit Optimization Results. $S_{\mathrm{cap}}^{(2)}$ is the
% stabilizer-weighted capability score (max\,=\,16{,}340); $S_{\mathrm{qual}}^{(2)}$
% is the quality score. Mean $G_{2Q}$$\downarrow$ is measured only among successful
% circuits. Gemini was evaluated at 1 attempt only (see text).}
% \label{tab:b2}
% \resizebox{\columnwidth}{!}{%
% \begin{tabular}{llcccc}
% \toprule
% Model & Config & Success & $S_{\mathrm{cap}}^{(2)}$ scores & $S_{\mathrm{qual}}^{(2)}$ scores & Mean $G_{2Q}$$\downarrow$ \\
% \midrule
% Claude   & 1 att / ---   & 26.0\% & 4{,}028  &   308.1  & 16.3\% \\
% Claude  & 15 att / 300s & 35.4\% & 3{,}162  &   207.6  & 15.2\% \\
% Claude  & 15 att / 900s & 72.9\% & 9{,}186  & 2{,}910.7 & 66.4\% \\
% \midrule
% GPT          & 1 att / ---   & 28.1\% & 4{,}912  &   147.5  & 10.3\% \\
% GPT          & 15 att / 300s & 42.7\% & 4{,}154  &   230.3  & 15.1\% \\
% GPT         & 15 att / 900s & 71.9\% & 8{,}994  & 2{,}756.9 & 68.0\% \\
% \midrule
% Gemini  & 1 att / ---   & 65.1\% & 7{,}738  & 2{,}791.9 & 65.3\% \\
% \bottomrule
% \multicolumn{6}{l}{$^{*}$Gemini evaluated at 1 attempt only; see Section~\ref{sec:results-b2}.}
% \end{tabular}}
% \end{table}

\begin{table}[t]
\centering
\caption{B2 Circuit Optimization Results. Summary of capability and quality
scores across models and configurations. Mean $G_{2Q}{\downarrow}$ is
measured only among successful circuits. The configuration is listed
as attempts/timeout in seconds.}
\label{tab:b2}
\resizebox{\columnwidth}{!}{%
\begin{tabular}{llcccc}
\toprule
Model & Config & Success & $S_{\mathrm{cap}}^{(2)}$ & $S_{\mathrm{qual}}^{(2)}$ & Mean $G_{2Q}\downarrow$ \\
\midrule
Claude & 1 att / 900s   & 26.0\% & 4{,}028 & 308   & 16.3\% \\
Claude & 15 att / 300s & 35.4\% & 3{,}162 & 208   & 15.2\% \\
Claude & 15 att / 900s & 72.9\% & 9{,}186 & 2{,}911 & 66.4\% \\
\midrule
GPT    & 1 att / 900s   & 28.1\% & 4{,}912 & 148   & 10.3\% \\
GPT    & 15 att / 300s & 42.7\% & 4{,}154 & 230   & 15.1\% \\
GPT    & 15 att / 900s & 71.9\% & 8{,}994 & 2{,}757 & 68.0\% \\
\midrule
Gemini & 1 att / 900s   & 65.1\% & 7{,}738 & 2{,}792 & 65.3\% \\
\bottomrule
\multicolumn{6}{l}{$^{*}$Gemini evaluated at 1 attempt only; see Section~\ref{sec:results-b2}.}
\end{tabular}}
\end{table}

At their best configurations, Claude and GPT achieve comparable capability
scores ($S_{\mathrm{cap}}^{(2)}$ of 9{,}186 and 8{,}994 out of 16{,}340),
meaning both agents successfully optimize roughly half the benchmark's
stabilizer-weighted difficulty. However, both quality scores
($S_{\mathrm{qual}}^{(2)}$ of 2{,}911 and 2{,}757) fall well below the
capability scores, reflecting that successful optimizations typically
capture only a fraction of the available improvement. The gap between
capability and quality scores widens on larger codes, where agents are more
likely to find a marginal improvement than deeper ones.

The quality scores are driven primarily by two-qubit gate reduction:
at the best configuration, successful circuits achieve mean
$G_{2Q}\downarrow$ of 66\% (Claude) and 68\% (GPT), with 96--98\% of
successful circuits reducing $G_{2Q}$. The majority also reduce depth
simultaneously; the remaining cases improve $G_{2Q}$ alone, typically on
circuits where a single gate cancellation does not cascade to a shorter
critical path.

Timeout budget again proves decisive: extending from 300\,s to 900\,s at
15 attempts more than doubles $S_{\mathrm{cap}}^{(2)}$ for both Claude
(3{,}162\,$\to$\,9{,}186) and GPT (4{,}154\,$\to$\,8{,}994), with quality
scores increasing by an order of magnitude.

Figure~\ref{fig:difficulty-curves} shows the cumulative capability
score $S_{\mathrm{cap}}^{(2)}$ as a function of stabilizer count at the
best configuration per agent. All models degrade sharply with circuit
complexity: Claude optimizes 98\% of small circuits (${\leq}38$ stabilizers)
but only 36\% of extra-large ones ($>$132 stabilizers); GPT similarly drops
from 94\% to 26\%, validating that the dataset presents a meaningful
difficulty spread.

Gemini Pro was only evaluated at $A\!=\!1$ and achieved
$S_{\mathrm{cap}}^{(2)}$ of 7{,}738 and $S_{\mathrm{qual}}^{(2)}$ of
2{,}792 in a single attempt---comparable to what Claude and GPT require
15 attempts and 900\,s to reach, suggesting stronger single-shot
optimization capability. Additional configurations were not tested because
during multi-attempt runs the agent deleted its own scratch files, causing
the evaluation harness to terminate early. A controlled re-evaluation
under a sandboxed harness is left to future work.

%% ============================================================
%%  IV-C. B3 Results: Fault Tolerance (0.75 pages)
%%  Writer: Sylvie | Editor: Cordelia
%% ============================================================
\subsection{B3 Results: Fault Tolerance}
\label{sec:results-b3}
The use of Claude, Gemini, and GPT to transform non-fault-tolerant circuits into fault-tolerant circuits was largely unsuccessful. The agents were tested under four different configurations in an effort to improve successful fault-tolerant circuit generation. The configurations are summarized as follows:
\begin{enumerate}
    \item Prompt~$a$, 15 attempts, 300 second timeout;
    \item Prompt~$b$, 15 attempts, 300 second timeout;
    \item Prompt~$b$, 15 attempts, 900 second timeout;
    \item Prompt~$b$, 1 attempt, 900 second timeout.
\end{enumerate}
In the first configuration, the agents were provided with Prompt~$a$, which includes a detailed description of the task, definitions of fault tolerance and flag qubits, transformation guidelines, and output rules. In subsequent configurations, Prompt~$b$ was used; it is largely identical to Prompt~$a$ but additionally includes a description of the fault tolerance score and its associated equations.

The cumulative capability score for fault tolerance can be found in Figure~\ref{fig:difficulty-curves}. Under the first, most effective configuration, Claude achieved a capability score of 2{,}830, Gemini achieved 276, and GPT achieved 2{,}220, out of a maximum possible score of 16{,}340. The first configuration consistently outperformed the others, which generally produced substantially lower scores (Table~\ref{tab:ft_combined_results}). This is consistent with Benchmark~B1, where the more complex prompt resulted in lower capability scores compared to the less verbose formulation.

Model-specific circuit generation strategies varied across configurations but did not appear to be significantly influenced by the choice of prompt. The observed circuit generation strategies are summarized as follows:
\begin{enumerate}
    \item \textbf{Claude:} Generally favored adding five or fewer flag qubits during circuit generation.
    
    \item \textbf{Gemini:} Consistently avoided adding five or fewer qubits across all configurations. When allocated 900~s per circuit, it predominantly added more than five qubits, whereas under a 300~s constraint it more frequently left circuits unchanged.
    
    \item \textbf{GPT:} Exhibited less consistent behavior across configurations, with each configuration demonstrating a different flag-qubit allocation strategy.
\end{enumerate}

Across all configurations, the evaluated agents exhibited limited success in generating fault-tolerant circuits that preserved the complete set of input stabilizers.

\begin{table}[t]
\centering
\caption{B3 Fault Tolerant circuit generation results. Summary of fault tolerance (FT), capability, and quality scores across models and configurations. The configuration is listed as prompt/attempts/timeout in seconds.}
\label{tab:ft_combined_results}
\resizebox{\columnwidth}{!}{%
\begin{tabular}{llcccc}
\toprule
Model & Config & Avg.\ of all $\mathrm{FT}>0$ & \# circuits w/ $\mathrm{FT}>0$ & $S_{\mathrm{cap}}^{(3)}$ & $S_{\mathrm{qual}}^{(3)}$ \\
\midrule
Claude & $a$ / 15 att / 300s & 0.119 & 152 & 2{,}830 & 253 \\
Claude & $b$ / 15 att / 300s & 0.116 & 80  & 1{,}018 & 83 \\
Claude & $b$ / 15 att / 900s & 0.161 & 71  & 1{,}994 & 55 \\
Claude & $b$ / 1 att / 900s  & 0.073 & 60  & 762     & 102 \\
\midrule
Gemini & $a$ / 15 att / 300s & 0.175 & 82  & 276     & 182 \\
Gemini & $b$ / 15 att / 300s & 0.242 & 51  & 154     & 78 \\
Gemini & $b$ / 15 att / 900s & 0.499 & 55  & 1{,}226 & 656 \\
Gemini & $b$ / 1 att / 900s  & 0.725 & 25  & 546     & 416 \\
\midrule
GPT    & $a$ / 15 att / 300s & 0.095 & 151 & 2{,}220 & 211 \\
GPT    & $b$ / 15 att / 300s & 0.102 & 76  & 878     & 51 \\
GPT    & $b$ / 15 att / 900s & 0.180 & 69  & 2{,}130 & 129 \\
GPT    & $b$ / 1 att / 900s  & 0.059 & 39  & 658     & 89 \\
\bottomrule
\end{tabular}}
\end{table}

\section{Limitations and Future Work}
\label{sec:discussion}

The 168 tensor product circuits are derived from only 24 base codes,
meaning structural patterns repeat across the dataset. A model that
learns to solve a base code may partially generalize to its tensor
products without reasoning about the combined code, which could inflate
scores relative to performance on genuinely novel codes. Additionally,
the current base code set does not cover all practically relevant QEC
families; hypergraph product codes~\cite{tillich2014hypergraph}, fiber
bundle codes~\cite{hastings2021fiber}, and other recent quantum LDPC
constructions beyond bivariate bicycle
codes~\cite{panteleev2022asymptotically} are absent from the benchmark.

The benchmark is restricted to stabilizer circuits over Clifford gates.
Non-Clifford operations such as the $T$ gate are not covered, and results
may not generalize to universal quantum circuit synthesis.

Circuits are expressed in Stim's text format, which may advantage models
with Stim exposure in training data. Extending to OpenQASM or other
formats would broaden applicability.

Several directions for future work emerge naturally. The dataset can be
expanded with hypergraph product codes, fiber bundle codes, and syndrome
extraction circuits; compositional decomposition---breaking large circuits
into verified subcircuits---is a natural next task beyond B3. The
structured, verifiable nature of stabilizer circuits also makes
\textsc{StabilizerBench} well-suited as a training signal for fine-tuning
or reinforcement learning from oracle feedback, opening a path toward
specialized quantum circuit agents. A public leaderboard with versioned
dataset snapshots and automated submission would allow the community to
track progress as new model generations are released, following the model
established by SWE-bench~\cite{jimenez2024swebench}.

% Sarju: write this section.
% Content guidance (from project plan):
% - Benchmark quality analysis against the six design criteria (Rohe et al.)
% - Limitations & threats to validity:
%   * Limited to stabilizer circuits (may not generalize to arbitrary quantum programs)
%   * Three models from one point in time; capabilities evolving rapidly
%   * Prompt sensitivity — different prompts could yield different results
%   * Stim-specific circuit format
% - Extensibility:
%   * New code families, new tasks (syndrome extraction, compositional decomposition)
%   * Fine-tuning LLMs on quantum circuit data
%   * Community use: open-source plan, leaderboard vision

%% ============================================================
%%  SECTION VI — CONCLUSION (~0.25 pages)
%%  Writer: Andres | Editor: Sylvie
%% ============================================================
\section{Conclusion}
\label{sec:conclusion}

We introduced \textsc{StabilizerBench}, a benchmark of 192 stabilizer codes and three tasks (state-preparation generation, circuit optimization, and fault-tolerant synthesis) with polynomial-time verification oracles and a unified scoring framework.
Baseline results from three frontier models show the benchmark is discriminative: agents solve most state-preparation instances but struggle with optimization quality and fault tolerance, where capability scores remain below 3,000 in all of the configurations.
A recurring practical finding is that performance is sensitive to configuration: more time per attempt helps more than increased attempts, and minimal prompts outperform detailed ones. This highlights the value of a flexible harness that lets users tune prompts, timeouts, attempt budgets, and agent strategies to find what works best for their model.
The benchmark, dataset, and evaluation harness are publicly available\footnote{\url{https://github.com/uw-math-ai/quantum-ai}} and designed to accommodate new code families, models, and circuit formats. 

\section*{Acknowledgments}
This research was developed as part of the University of Washington Math AI Lab.

%% ============================================================
%%  REFERENCES
%% ============================================================
\bibliographystyle{IEEEtran} 
\bibliography{Mybib.bib}

@inproceedings{shor1997,
  title={Polynomial-Time Algorithms for Prime Factorization and Discrete Logarithms on a Quantum Computer},
  author={Shor, Peter W.},
  journal={SIAM Journal on Computing},
  volume={26},
  number={5},
  pages={1484--1509},
  year={1997},
  doi={10.1137/S0097539795293172}
}

@article{aspuru2005,
  title={Simulated Quantum Computation of Molecular Energies},
  author={Aspuru-Guzik, Al{\'a}n and Dutoi, Anthony D. and Love, Peter J. and Head-Gordon, Martin},
  journal={Science},
  volume={309},
  number={5741},
  pages={1704--1707},
  year={2005},
  doi={10.1126/science.1113479}
}

@book{nielsen2010quantum,
  title={Quantum Computation and Quantum Information: 10th Anniversary Edition},
  author={Nielsen, Michael A. and Chuang, Isaac L.},
  year={2010},
  publisher={Cambridge University Press},
  doi={10.1017/CBO9780511976667}
}

@article{gottesman1998heisenberg,
  title={The {H}eisenberg Representation of Quantum Computers},
  author={Gottesman, Daniel},
  journal={arXiv preprint quant-ph/9807006},
  year={1998}
}

@article{aaronson2004improved,
  title={Improved Simulation of Stabilizer Circuits},
  author={Aaronson, Scott and Gottesman, Daniel},
  journal={Physical Review A},
  volume={70},
  number={5},
  pages={052328},
  year={2004},
  publisher={APS},
  doi={10.1103/PhysRevA.70.052328}
}

@article{knill1997theory,
  title={Theory of Quantum Error-Correcting Codes},
  author={Knill, Emanuel and Laflamme, Raymond},
  journal={Physical Review A},
  volume={55},
  number={2},
  pages={900},
  year={1997},
  publisher={APS},
  doi={10.1103/PhysRevA.55.900}
}

@article{terhal2015quantum,
  title={Quantum Error Correction for Quantum Memories},
  author={Terhal, Barbara M.},
  journal={Reviews of Modern Physics},
  volume={87},
  number={2},
  pages={307},
  year={2015},
  publisher={APS},
  doi={10.1103/RevModPhys.87.307}
}

@article{cleve1997efficient,
  title={Efficient Computations of Encodings for Quantum Error Correction},
  author={Cleve, Richard and Gottesman, Daniel},
  journal={Physical Review A},
  volume={56},
  number={1},
  pages={76},
  year={1997},
  publisher={APS},
  doi={10.1103/PhysRevA.56.76}
}

@article{Prabhu_2023,
  title={Fault-tolerant syndrome extraction and cat state preparation with fewer qubits},
  volume={7},
  ISSN={2521-327X},
  url={http://dx.doi.org/10.22331/q-2023-10-24-1154},
  DOI={10.22331/q-2023-10-24-1154},
  journal={Quantum},
  publisher={Verein zur Forderung des Open Access Publizierens in den Quantenwissenschaften},
  author={Prabhu, Prithviraj and Reichardt, Ben W.},
  year={2023},
  month=oct,
  pages={1154}
}

@phdthesis{Gottesman1997,
  author={Daniel Gottesman},
  title={Stabilizer Codes and Quantum Error Correction},
  school={California Institute of Technology},
  address={Pasadena, California},
  year={1997},
  eprint={quant-ph/9705052},
  archivePrefix={arXiv}
}

@article{cross2022openqasm3,
  title={{OpenQASM} 3: A Broader and Deeper Quantum Assembly Language},
  author={Cross, Andrew W. and Javadi-Abhari, Ali and Alexander, Thomas and de Beaudrap, Niel and Bishop, Lev S. and Heidel, Steven and Ryan, Colm A. and Sivarajah, Prashant and Smolin, John and Gambetta, Jay M. and Johnson, Blake R.},
  journal={ACM Transactions on Quantum Computing},
  volume={3},
  number={3},
  pages={1--50},
  year={2022},
  publisher={ACM},
  doi={10.1145/3505636}
}

@article{svore2018qsharp,
  title={{Q\#}: Enabling Scalable Quantum Computing and Development with a High-Level {DSL}},
  author={Svore, Krysta M. and Geller, Alan and Troyer, Matthias and Azariah, John and Granade, Christopher and Heim, Bettina and Kliuchnikov, Vadym and Mykhailova, Mariia and Paz, Andres and Roetteler, Martin},
  journal={Proceedings of the Real World Domain Specific Languages Workshop (RWDSL)},
  year={2018},
  publisher={ACM},
  doi={10.1145/3183895.3183901}
}

@misc{guppy2024,
  title={Guppy: A Pythonic Quantum Programming Language},
  author={{Quantinuum}},
  year={2024},
  howpublished={\url{https://github.com/CQCL/guppylang}},
  note={Accessed: 2026-03-01}
}

@inproceedings{voichick2023qunity,
  title={Qunity: A Unified Language for Quantum and Classical Computing},
  author={Voichick, Finn and Li, Liyi and Rand, Robert and Hicks, Michael},
  booktitle={Proceedings of the ACM on Programming Languages (POPL)},
  volume={7},
  year={2023},
  doi={10.1145/3571225}
}

@article{yuan2024tower,
  title={Tower: Data Representations in a Quantum Programming Language},
  author={Yuan, Charles and Carbin, Michael},
  journal={Proceedings of the ACM on Programming Languages (POPL)},
  volume={8},
  year={2024},
  doi={10.1145/3632900}
}

@article{jimenez2024swebench,
  title={{SWE}-bench: Can Language Models Resolve Real-World {GitHub} Issues?},
  author={Jimenez, Carlos E. and Yang, John and Wettig, Alexander and Yao, Shunyu and Pei, Kexin and Press, Ofir and Narasimhan, Karthik},
  journal={arXiv preprint arXiv:2310.06770},
  year={2024}
}

@article{chen2021humaneval,
  title={Evaluating Large Language Models Trained on Code},
  author={Chen, Mark and Tworek, Jerry and Jun, Heewoo and Yuan, Qiming and {de Oliveira Pinto}, Henrique Pond{\'e} and Kaplan, Jared and Edwards, Harri and Burda, Yuri and Joseph, Nicholas and Brockman, Greg and others},
  journal={arXiv preprint arXiv:2107.03374},
  year={2021},
  note={Introduces the HumanEval benchmark}
}

@article{austin2021mbpp,
  title={Program Synthesis with Large Language Models},
  author={Austin, Jacob and Odena, Augustus and Nye, Maxwell and Bosma, Maarten and Michalewski, Henryk and Dohan, David and Jiang, Ellen and Cai, Carrie and Terry, Michael and Le, Quoc and Sutton, Charles},
  journal={arXiv preprint arXiv:2108.07732},
  year={2021}
}

@article{gidney2021stim,
  title={Stim: A Fast Stabilizer Circuit Simulator},
  author={Gidney, Craig},
  journal={Quantum},
  volume={5},
  pages={497},
  year={2021},
  doi={10.22331/q-2021-07-06-497}
}

@article{rohe2025quantum,
  title={Quantum Computer Benchmarking: An Explorative Systematic Literature Review},
  author={Rohe, Tobias and Harjes Ruiloba, Federico and Egger, Sabrina and von Beck, Sebastian and Stein, Jonas and Linnhoff-Popien, Claudia},
  journal={arXiv preprint arXiv:2509.03078},
  year={2025}
}

@article{Peham_2025,
  title={Automated Synthesis of Fault-Tolerant State Preparation Circuits for Quantum Error-Correction Codes},
  volume={6},
  ISSN={2691-3399},
  url={http://dx.doi.org/10.1103/PRXQuantum.6.020330},
  DOI={10.1103/prxquantum.6.020330},
  number={2},
  journal={PRX Quantum},
  publisher={American Physical Society (APS)},
  author={Peham, Tom and Schmid, Ludwig and Berent, Lucas and M{\"u}ller, Markus and Wille, Robert},
  year={2025},
  month=may
}

@inproceedings{huang2019assertions,
  title={Statistical Assertions for Validating Patterns and Finding Bugs in Quantum Programs},
  author={Huang, Yipeng and Martonosi, Margaret},
  booktitle={Proceedings of the 46th International Symposium on Computer Architecture (ISCA)},
  year={2019},
  doi={10.1145/3307650.3322213}
}

@article{luo2022bugs,
  title={A Comprehensive Study of Bug Fixes in Quantum Programs},
  author={Luo, Junjie and Zhao, Pengzhan and Miao, Zhongtao and Lan, Shuhan and Zhao, Jianjun},
  journal={arXiv preprint arXiv:2201.08662},
  year={2022}
}

@article{ramalho2024debugging,
  title={Testing and Debugging Quantum Programs: The Road to 2030},
  author={Ramalho, Neilson Carlos Leite and de Souza, Higor Amario and Chaim, Marcos Lordello},
  journal={arXiv preprint arXiv:2405.09178},
  year={2024}
}

@article{ma2025qmon,
  title={{QMon}: Monitoring the Execution of Quantum Circuits with Mid-Circuit Measurement and Reset},
  author={Ma, Ning and Zhao, Jianjun and Khomh, Foutse and Ali, Shaukat and Li, Heng},
  journal={arXiv preprint arXiv:2512.13422},
  year={2025}
}

@article{qiskit_humaneval,
  title={{Qiskit HumanEval}: An Evaluation Benchmark for Quantum Code Generative Models},
  author={Vishwakarma, Sanjay and Harkins, Francis and Golecha, Siddharth and Bajpe, Vishal Sharathchandra and Dupuis, Nicolas and Buratti, Luca and Kremer, David and Mezzacapo, Antonio and Tacchino, Francesco},
  journal={arXiv preprint arXiv:2406.14712},
  year={2024}
}

@article{quanbench,
  title={{QuanBench}: Benchmarking Quantum Code Generation with Large Language Models},
  author={Guo, Xiaoyu and Wang, Minggu and Zhao, Jianjun},
  journal={arXiv preprint arXiv:2510.16779},
  year={2025},
  note={Accepted at ASE 2025}
}

@article{qhackbench,
  title={{QHackBench}: Benchmarking Large Language Models for Quantum Code Generation Using {PennyLane} Hackathon Challenges},
  author={Basit, Abdul and Shao, Minghao and Asif, Muhammad Haider and Innan, Nouhaila and Kashif, Muhammad and Marchisio, Alberto and Shafique, Muhammad},
  journal={arXiv preprint arXiv:2506.20008},
  year={2025},
  note={To appear at IEEE QAI 2025}
}

@article{qcoder,
  title={{QCoder} Benchmark: Bridging Language Generation and Quantum Hardware through Simulator-Based Feedback},
  author={Mikuriya, Taku and Ishigaki, Tatsuya and Kawarada, Masayuki and Minami, Shunya and Kadowaki, Tadashi and Suzuki, Yohichi and Naito, Soshun and Takata, Shunya and Kato, Takumi and Basseda, Tamotsu and Yamada, Reo and Takamura, Hiroya},
  journal={arXiv preprint arXiv:2510.26101},
  year={2025},
  note={Accepted at INLG 2025}
}

@article{PhysRevLett.121.050502,
  title={Quantum Error Correction with Only Two Extra Qubits},
  author={Chao, Rui and Reichardt, Ben W.},
  journal={Phys. Rev. Lett.},
  volume={121},
  issue={5},
  pages={050502},
  numpages={5},
  year={2018},
  month={Aug},
  publisher={American Physical Society},
  doi={10.1103/PhysRevLett.121.050502},
  url={https://link.aps.org/doi/10.1103/PhysRevLett.121.050502}
}

@article{Chamberland2018flagfaulttolerant,
  doi={10.22331/q-2018-02-08-53},
  url={https://doi.org/10.22331/q-2018-02-08-53},
  title={Flag fault-tolerant error correction with arbitrary distance codes},
  author={Chamberland, Christopher and Beverland, Michael E.},
  journal={{Quantum}},
  issn={2521-327X},
  publisher={{Verein zur F{\"{o}}rderung des Open Access Publizierens in den Quantenwissenschaften}},
  volume={2},
  pages={53},
  month=feb,
  year={2018}
}

@article{tomita2014rotated,
  title={Low-Distance Surface Codes under Realistic Quantum Noise},
  author={Tomita, Yu and Svore, Krysta M.},
  journal={Physical Review A},
  volume={90},
  number={6},
  pages={062320},
  year={2014},
  publisher={APS},
  doi={10.1103/PhysRevA.90.062320}
}

@article{bombin2006topological,
  title={Topological Quantum Distillation},
  author={Bombin, H. and Martin-Delgado, M. A.},
  journal={Physical Review Letters},
  volume={97},
  number={18},
  pages={180501},
  year={2006},
  publisher={APS},
  doi={10.1103/PhysRevLett.97.180501}
}

@article{landahl2011fault,
  title={Fault-Tolerant Quantum Computing with Color Codes},
  author={Landahl, Andrew J. and Anderson, Jonas T. and Rice, Patrick R.},
  journal={arXiv preprint arXiv:1108.5738},
  year={2011}
}

@article{self2024iceberg,
   title={Protecting expressive circuits with a quantum error detection code},
   volume={20},
   ISSN={1745-2481},
   url={http://dx.doi.org/10.1038/s41567-023-02282-2},
   DOI={10.1038/s41567-023-02282-2},
   number={2},
   journal={Nature Physics},
   publisher={Springer Science and Business Media LLC},
   author={Self, Chris N. and Benedetti, Marcello and Amaro, David},
   year={2024},
   month=jan, pages={219–224} }

@article{goto2024hypercube,
  title={High-Performance Fault-Tolerant Quantum Computing with Many-Hypercube Codes},
  author={Goto, Hayato},
  journal={Science Advances},
  volume={10},
  pages={eadp6388},
  year={2024},
  doi={10.1126/sciadv.adp6388}
}

@article{bravyi2024high,
  title={High-Threshold and Low-Overhead Fault-Tolerant Quantum Memory},
  author={Bravyi, Sergey and Cross, Andrew W. and Gambetta, Jay M. and Maslov, Dmitri and Rall, Patrick and Yoder, Theodore J.},
  journal={Nature},
  volume={627},
  pages={778--782},
  year={2024},
  doi={10.1038/s41586-024-07107-7}
}

@article{laflamme1996perfect,
  title={Perfect Quantum Error Correcting Code},
  author={Laflamme, Raymond and Miquel, Cesar and Paz, Juan Pablo and Zurek, Wojciech Hubert},
  journal={Physical Review Letters},
  volume={77},
  number={1},
  pages={198},
  year={1996},
  publisher={APS},
  doi={10.1103/PhysRevLett.77.198}
}

@article{steane1996multiple,
  title={Multiple-Particle Interference and Quantum Error Correction},
  author={Steane, Andrew M.},
  journal={Proceedings of the Royal Society A},
  volume={452},
  number={1954},
  pages={2551--2577},
  year={1996},
  doi={10.1098/rspa.1996.0136}
}

@article{shor1995scheme,
  title={Scheme for Reducing Decoherence in Quantum Computer Memory},
  author={Shor, Peter W.},
  journal={Physical Review A},
  volume={52},
  number={4},
  pages={R2493},
  year={1995},
  publisher={APS},
  doi={10.1103/PhysRevA.52.R2493}
}

@article{steane1996reed_muller,
  title={Quantum {Reed-Muller} Codes},
  author={Steane, Andrew M.},
  journal={IEEE Transactions on Information Theory},
  volume={45},
  number={5},
  pages={1701--1703},
  year={1999},
  doi={10.1109/18.771249}
}

@article{knill2005realistic,
  title={Quantum Computing with Realistically Noisy Devices},
  author={Knill, Emanuel},
  journal={Nature},
  volume={434},
  pages={39--44},
  year={2005},
  doi={10.1038/nature03350}
}

@article{paetznick2024logical,
  title={Demonstration of Logical Qubits and Repeated Error Correction with Better-Than-Physical Error Rates},
  author={Paetznick, Adam and da Silva, Marcus P. and Ryan-Anderson, Ciaran and Bello-Rivas, Juan M. and Camara, Charles H. and Craft, Jonathan and Dalzell, Alexander M. and Eickbusch, Alec and Gidney, Craig and Graydon, Matthew and others},
  journal={arXiv preprint arXiv:2404.02280},
  year={2024}
}

@article{tillich2014hypergraph,
  title={Quantum {LDPC} Codes with Positive Rate and Minimum Distance Proportional to the Square Root of the Blocklength},
  author={Tillich, Jean-Pierre and Z{\'e}mor, Gilles},
  journal={IEEE Transactions on Information Theory},
  volume={60},
  number={2},
  pages={1193--1202},
  year={2014},
  doi={10.1109/TIT.2013.2292061}
}

@inproceedings{hastings2021fiber,
  title={Fiber Bundle Codes: Breaking the $N^{1/2}\,\mathrm{polylog}(N)$ Barrier for Quantum {LDPC} Codes},
  author={Hastings, Matthew B. and Haah, Jeongwan and O'Donnell, Ryan},
  booktitle={Proceedings of the 53rd Annual ACM Symposium on Theory of Computing (STOC)},
  pages={1276--1288},
  year={2021},
  doi={10.1145/3406325.3451005}
}

@inproceedings{panteleev2022asymptotically,
  title={Asymptotically Good Quantum and Locally Testable Classical {LDPC} Codes},
  author={Panteleev, Pavel and Kalachev, Gleb},
  booktitle={Proceedings of the 54th Annual ACM Symposium on Theory of Computing (STOC)},
  pages={375--388},
  year={2022},
  doi={10.1145/3519935.3520017}
}

\end{document}